\documentclass[11pt,preprint,aps,nofootinbib,shownopacs]{revtex4}
\usepackage{graphics}
\usepackage{psfig}
\usepackage{rotating}
\usepackage{graphicx}

\newcommand*{\be}{\begin{equation}}
\newcommand*{\ee}{\end{equation}}
\newcommand*{\bea}{\begin{eqnarray}}
\newcommand*{\eea}{\end{eqnarray}}
\providecommand*{\ler}{\stackrel{\scriptstyle <}{\scriptstyle \sim}}
\providecommand*{\ger}{\stackrel{\scriptstyle >}{\scriptstyle \sim}}
\newcommand{\nn}{\nonumber}
\newcommand{\Frac}[2]{\frac{\displaystyle{#1}}{\displaystyle{#2}}}
\newcommand{\lsim}{\raise0.3ex\hbox{$\;<$\kern-0.75em\raise-1.1ex\hbox{$\sim\;$}}}
\newcommand{\gsim}{\raise0.3ex\hbox{$\;>$\kern-0.75em\raise-1.1ex\hbox{$\sim\;$}}}

\begin{document}
\title{Massive Neutrinos and Flavour Violation}
\author{Antonio Masiero}
\email{masiero@pd.infn.it}
\affiliation{Dip. di Fisica `Galileo Galilei', Univ. di Padova, and
INFN, Sezione di Padova, via F. Marzolo 8,  I-35131, Padova, Italy} 
\author{Sudhir K. Vempati}
\email{vempati@pd.infn.it}
\affiliation{Dip. di Fisica `Galileo Galilei', Univ. di Padova, and
INFN, Sezione di Padova, via F. Marzolo 8,  I-35131, Padova, Italy} 
\author{Oscar Vives}
\email{oscar.vives@cern.ch}
\affiliation{Theory Group, Physics Department, CERN, Geneva, Switzerland\\}
\preprint{CERN-PH-TH/2004-142}

\begin{abstract}
\vskip 0.5cm
In spite of the large lepton flavour violation (LFV) observed in neutrino 
oscillations, within the Standard Model, we do \textit{not} expect any visible 
LFV in the charged lepton sector ($\mu \to e, \gamma$, $\tau \to \mu, \gamma$, 
etc.). On the contrary, the presence of new physics close to the electroweak
scale can enhance the amplitudes of these processes. We discuss
this in general and focus on a particularly interesting case: the marriage of
low-energy supersymmetry (SUSY) and seesaw mechanism for neutrino masses (SUSY seesaw). 
Several ideas presented in this context are reviewed both in the bottom-up and
top-down approaches. We show that there exist attractive models where the rate
for LFV processes can attain values to be probed in pre-LHC experiments. 
\end{abstract}

\maketitle
\section{Introduction}

Since the last couple of years of the previous century, tremendous progress 
has been made in our understanding of the nature of the most 
elusive Standard Model (SM) particles, the neutrinos. The main message of
various experimental results has been that neutrinos have non-standard
properties: they have masses and their flavour states mix and, indeed, 
with very large mixings.
In the Standard Model, the phenomenon of flavour mixing is not surprising,
as its presence has already been well established in the quark sector. 
Furthermore, the induced phenomenological implications such as 
$K^0$-$\bar{K^0}$ oscillations, $B_d$-$\bar{B_d}$ mixing and  $b \to s, \gamma$ 
have been well understood as well as measured with high 
precision. Given this, and non-zero neutrino flavour
mixing, one would expect that similar phenomenological and theoretical 
implications can now emanate from the leptonic sector. 

The more obvious of the phenomenological implications of neutrino flavour 
mixing phenomena are processes like neutrino oscillations. 
On the other hand, the presence of this neutrino flavour mixing would also 
induce flavour mixing in its Isodoublet partners, the charged leptons. 
This mixing, induced at the 1-loop level through gauge bosons, is manifested by
rare decay processes such as $\mu \to e, \gamma$, $\tau \to \mu, \gamma$, etc. 
If only the neutrinos carry this information on flavour mixing, as in the
Standard Model with massive neutrinos, these processes are expected to 
be proportional to the ratio of masses of neutrinos over the masses of
the W bosons, leading to extremely tiny branching ratios.

This situation can dramatically change if there are some new additional
particles carrying lepton flavour numbers, which have very large masses
and simultaneously mix among themselves. The presence of such particles
could lead to enhanced branching ratios for the above processes, perhaps 
bringing them into the realm of observability of the present and next 
generations of experiments. On the other hand, non-observability of these
processes can lead to strong constraints on the nature of new physics,
which is expected to be present just above the electroweak scale. Such
constraints already exist from the hadronic sector, albeit riddled
with typical uncertainties associated with them. The leptonic flavour changing 
processes, on the other hand, do not suffer from these uncertainties,
and thus lead to stronger constraint on any new flavour violating physics. 
In fact, the present constraint on BR($\mu \to e, \gamma$) has long been 
considered to be the most stringent constraint on any new flavour physics. 
To get a feeling as to  where we stand, we provide here a list of present 
and upcoming experimental limits: 

\begin{center} 
\textit{Present limits}\\[12pt] 
\begin{tabular}{rclc}
$BR(\mu \to e \gamma)$ &$\leq$& $1.2\times 10^{-11}$ & \cite{mega}\\
 $BR(\tau \to \mu \gamma)$ &$\leq$& $3.1\times 10^{-7}$ & \cite{belletmg}\\ 
$BR(\tau \to e \gamma)$ &$\leq$& $3.7\times 10^{-7}$ & \cite{belletalk} 
\end{tabular} 
\end{center}

\begin{center}
\textit{Upcoming limits}\\[12pt]
\begin{tabular}{rclc}
$BR(\mu \to e \gamma)$ &$\leq$& $10^{-13} $--$ 10^{-14}$ & \cite{psi}\\
$BR(\tau \to \mu \gamma)$ &$\leq$& $10^{-8}$ & \cite{belletalk}\\
$BR(\tau \to e \gamma)$ &$\leq$& $10^{-8}$ & \cite{belletalk}
\end{tabular}
\end{center} 

The impact of these limits could be felt in a wide class of new
physics models setting in at a scale close to the electroweak scale. A
particularly interesting class of models are the supersymmetric (SUSY)
Standard Model(s).  In these models, the supersymmetric partners of the
leptons, namely the sleptons, carry the same flavour quantum numbers
as the SM leptons.  Since they are expected to be heavy, around the
TeV scale, if flavour mixing is present in the (s)leptonic sector,
large branching ratios are expected for the afore-mentioned rare decay
processes. Interestingly enough, this naturally occurs if we marry
the idea of low energy supersymmetry to the mechanism of seesaw \cite{seesawrefs}
giving rise to small neutrino masses (SUSY seesaw \cite{fbam}). 
 In the present review, we will mainly concentrate on
this class of models.

\section{Supersymmetric Models and lepton flavour violation}
The study of flavour violation in supersymmetric models is certainly
quite complicated. Indeed, to the usual intricacies involved in the
FCNC computations in the SM, we add several new SUSY contributions. The
main source of this difficulty is the soft supersymmetry breaking
lagrangian, which can in general contain a large number of flavour
violating couplings, leading to significant constraints on its
parameters \cite{gabbiani,fcncreview}. An appealing manner to cope with these
tight constraints is to consider only a particular classes of soft
lagrangians, which result from models that break supersymmetry
in a flavour blind manner, as in mSUGRA, Anomaly mediated
supersymmetry breaking (AMSB) or its several variants, Gauge mediated
supersymmetry breaking (GMSB), etc \cite{kanekingphyrep}.  However, in
general, even after choosing a particular model of supersymmetry
breaking, flavour violation can still be present in the weak scale
lagrangian. The various sources of flavour violation in such a case
can be broadly classified as\footnote{We assume R-parity conservation
throughout the present work.}:\\
\noindent 
(i). In models of supersymmetry breaking based on supergravity or
superstring theories, although it is possible to achieve universality
or even no-scale boundary conditions under some assumptions on the
K{\"a}hler potential, non-universal soft terms are generically present in
the high scale effective lagrangian \cite{brignoleibanez}.\\
\noindent
(ii). In models with flavour symmetry imposed by a Froggatt--Nielsen
mechanism, flavour violating corrections to the soft potential could
be potentially large \cite{dudassavoy}. More so, if the flavon fields
contain SUSY breaking F-VEVs \cite{oscarross1,oscarross2}. \\
\noindent
(iii). Finally, the existence of new particles at high scales with
flavour violating couplings to the SM leptons (as right handed
neutrinos in a seesaw model \cite{fbam}) or the presence of new Yukawa
interactions (as in Grand Unified Theories where quark and leptonic
fields sit in the same (super)multiplets \cite{bhs}) can lead to
flavour violation at the weak scale. In this case, the flavour
violation is communicated to the low energy fields through
Renormalisation Group Equations (RGEs) \cite{hkr}.\\
\noindent
Obviously, this list is not exhaustive. There can be other exotic
sources which can be either additional heavy Higgs
particles \cite{kingpeddie} or are related to the localisation of the
fermions in higher dimensional space-time when MSSM/SUSY-GUT is
embedded in a extra-dimensional model \cite{hallnomura,choi}.

In the present review, we will concentrate on the flavour violation
solely due to a mechanism generating neutrino masses and
mixings. To this effect, we would consider models where the
\textit{strong} universality is assumed at the high scale and thus
point (i) would not be considered here. Similarly, we will not
consider either effects generated by the imposition of a flavour symmetry as
in (ii). Instead, the main aspect of this review will be to collect
some salient features of the flavour violation induced by a neutrino
mass model in a supersymmetric theory.  As was mentioned in the
introduction, the most natural and popular of them, the seesaw
mechanism, will be our main focus. Notice that these RGE effects are
always present in any SUSY seesaw model, independently of the presence
of the other sources in (i) and (ii). In this sense, the effects
considered here are independent of the particular mechanism of SUSY
breaking and mediation. These model dependent effects in (i) and (ii) 
can always be added to our results in the relevant cases.
 
\begin{figure}[ht]
\label{muegdiag}
\includegraphics[scale=0.80]{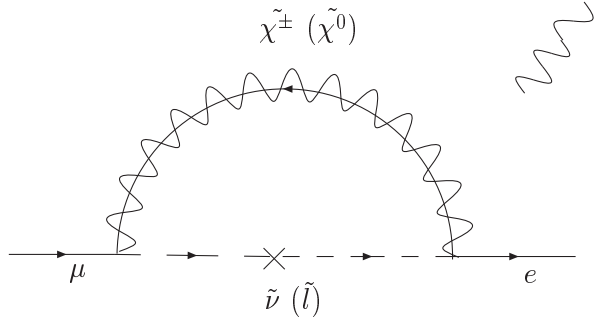}
\caption{The diagrams contributing to $\mu \to e, \gamma$ decays }
\end{figure}

Irrespective of the source, LFV at the weak scale can be parametrised in a 
model-independent manner in terms of a mass insertion (MI), $ \Delta^l_{ij}$, 
the flavour violating off-diagonal entry appearing in the slepton mass matrix\footnote{In
the basis where the charged lepton mass matrix is diagonal.}. 
These  MI are further subdivided into LL/LR/RL/RR types, labelled by the chirality 
of the corresponding SM fermions\footnote{$i,j,k$ denote generation indices throughout 
the present work.}. Depending on the model, one or several of these 
types of MI can simultaneously be present at the weak scale. In the presence of any of
these parameters, 1-loop diagrams mediated by gauginos, higgsinos (neutral and 3 
charged fermionic partners of gauge and Higgs bosons) and sleptons lead to lepton
flavour violating processes such as $\mu \to e + \gamma$, $\mu \to 3 e$, $\mu \to e$
conversion in nuclei, etc (an example diagram is shown in Fig.1). 
The strength of these processes crucially depends on the
MI factor $\delta^l_{ij} \equiv \Delta_{ij}^l / m_{\tilde{l}}^2$, where
$m_{\tilde{l}}^2$ is the average slepton mass. For $|\delta| < 1 $, which is expected to 
be the case for most models, one can always use the MI approximation \cite{hkr,amgab} to 
compute the amplitudes of the relevant processes. Such computations have been done 
long ago, considering the neutral gaugino diagrams \cite{fbam,gabbiani}. It has been
realised later that, in addition to the flavour violating LL/RR MI, considering the 
Higgsinos/gaugino mixing, as well as the flavour diagonal left-right mixing in the slepton 
mass matrix, can significantly enhance the amplitudes
of these processes at large tan $\beta$ \cite{largetanbeta}. These computations have since
then been updated by Hisano--Nomura \cite{HN} and Masina--Savoy \cite{MS}, including this
mixing as well as the charged gaugino/higgsino contribution\footnote{Another important feature is 
that the interference between various contributions could lead to suppressed amplitudes in some
regions of the parameter space \cite{hisano,HN,MS}. This typically occurs for RR type
MI as long as universality in the gaugino masses is maintained at the high scale.
Although in a completely generic situation without any universal boundary conditions, 
such cancellations can occur for LL type MI also \cite{stefanoyaguna1}.}. 
Taking the tan~$\beta$ factor into account, the branching ratio of $l_j \to l_i, \gamma$ for
the dominant LL MI is roughly given by: 
\be
\label{BR}
\mbox{BR} (l_j \to l_i \gamma) \approx \Frac{ \alpha^3 ~
~| \delta^{l}_{ij}|^2 }
{G_F^2~ m^4_{\rm SUSY}} \tan^2 \beta,
\ee
where $m_{\rm SUSY}$ represents the typical supersymmetry breaking mass such as
the gaugino/slepton mass. For large $|\delta| \sim 1$ or for many $\delta$'s 
present simultaneously, it is instructive to diagonalise the
slepton mass matrix and evaluate the precise amplitudes in the mass-eigenstate 
basis. A complete computation in this basis has been presented in \cite{hisano}
for several LFV processes such as $l_j \to l_i + \gamma$; $l_j \to 3 l_i$;~ $\mu \to
e $ conversion in nuclei. The processes discussed so far are the ones mediated 
by neutralino and chargino sector. However, Higgs bosons ($h^0,H^0,A^0$) are also 
sensitive to flavour violation and mediate processes such as $\mu \to e$ conversion
\cite{okada}, $\tau \to 3 \mu $ \cite{t3mu},
$\tau \to \mu \eta$ \cite{sher}. The amplitudes of these processes are sensitive to a 
higher degree in tan $\beta$ than the chargino/neutralino ones (the BRs grow as
$(\tan \beta)^6$, though they are suppressed by additional Yukawa couplings) and thus
could lead to large branching fractions at large tan $\beta$. Detailed studies on the
$\mu \to \tau$ sector have been presented in the literature \cite{andrea}.

In the rest of the review, we will consider the decay $l_j \to l_i, \gamma$ as the 
prototype signature of the lepton flavour violation. We will discuss several ideas
put forward in the literature on the sensitivity of this decay process in determining
the seesaw parameter space as well as the supersymmetric parameter space. Before this,
we will briefly review the SUSY seesaw mechanism and the generation of flavour violation
in this model. 

\section{Supersymmetric Seesaw and Leptonic Flavour Violation}

The seesaw mechanism  can be incorporated in the  
Minimal Supersymmetric Standard Model in a manner similar to what is  done in 
the Standard Model, by adding right-handed neutrino superfields to the
MSSM superpotential:
\be
\label{eq1}
W =  h^u_{ij} Q_i u_j^c H_2 + h^d_{ii} Q_i d_i^c H_1 + h^e_{ii} L_i e^c_i H_1 
+ h^\nu_{ij} L_i \nu^c_j H_2 
+ M_{R_{ii}} \nu_i^c \nu_i^c + \mu H_1 H_2 , 
\ee
where we are in the basis of diagonal charged lepton, down quark and right-handed
Majorana  mass matrices. 
$M_{R}$ represents the (heavy) Majorana mass matrix 
for the right-handed neutrinos.  Equation (\ref{eq1}) leads to the standard seesaw 
formula for the (light) neutrino mass matrix 
\begin{equation}
\label{seesaweq}
{\mathcal M}_\nu = - h^\nu M_{R}^{-1} h^{\nu~T} v_2^2,
\end{equation}
where $v_2$ is the vacuum expectation value (VEV) of the up-type
Higgs field, $H_2$. Under suitable conditions on $h^\nu$ and $M_R$,
the correct mass splittings and  mixing angles in $\mathcal{M}_\nu$ 
can be obtained. Detailed analyses deriving these conditions are 
already present in the literature \cite{seesawreviews}.  

Following the discussion in the
previous section, we will assume that the mechanism that breaks 
supersymmetry and conveys it to the observable sector at the high scale 
$\sim M_{\rm P}$ is flavour blind, as in mSUGRA. However, this flavour
blindness is not protected down to the weak scale \cite{fbam} \footnote{ This is 
always true in a gravity mediated supersymmetry breaking model, but it
also applies to other mechanisms under
some specific conditions \cite{yanagidagmsb,murayamaamsb}.}.
The slepton mass matrices are no longer invariant under RG evolution from the
super large scale where supersymmetry is mediated to the visible sector down to 
the seesaw scale, as the flavour violation present in the neutrino Dirac Yukawa
couplings $h^\nu$ is now `felt' by the slepton mass matrices in the presence of
heavy right-handed neutrinos. 

The weak-scale flavour violation so generated can be obtained
by solving the RGEs for the slepton mass matrices 
from the high scale to the scale of the right-handed neutrinos. Below 
this scale, the running of the FV slepton mass terms is RG-invariant 
as the right-handed neutrinos decouple from the theory. For the purpose of 
illustration, a leading log estimate can easily be obtained 
for these equations\footnote{Within mSUGRA, the leading log approximation works very well
for most of the parameter space, except for regions of large $M_{1/2}$ and low $m_0$. The
discrepancy with the exact result increases with low $\tan \beta$ \cite{petcovyag1}.}. 
Assuming the flavour blind mSUGRA specified by 
the high-scale parameters, $m_0$, the common scalar mass, $A_0$, 
the common trilinear coupling, and $M_{1/2}$, the universal gaugino mass, 
the flavour violating entries in these mass matrices at the weak scale 
are given as:

\be
\label{rgemi}
(\Delta^l_{ij})_{\rm LL} \approx -{3 m_0^2+A_0^2 \over 8 \pi^2} \sum_k
(h^\nu_{ik} h^{\nu *}_{jk}) \ln{M_{X} \over M_{R_k} },
\ee 
where $h^\nu$ are given in the basis of diagonal charged lepton masses
and diagonal Majorana right-handed neutrino mass matrix $M_R$, and
$M_X$ is the scale at which soft terms appear in the lagrangian. Given
this, the branching ratios for LFV rare decays $l_j \to l_i, \gamma$
can be roughly estimated using Eq.~(\ref{BR}).  From above it is
obvious that the amount of lepton flavour violation generated by the
SUSY seesaw at the weak scale crucially depends on the flavour
structure of $h^\nu$ and $M_R$, shown in Eq.~(\ref{eq1}), the `new'
sources of flavour violation not present in the MSSM.  If either the
neutrino Yukawa couplings or the flavour mixings present in $h^\nu$
are very tiny, the strength of LFV will be significantly reduced.
Further, if the right-handed neutrino masses were heavier than the
supersymmetry breaking scale (as in GMSB models) they would decouple
from the theory before the SUSY soft breaking matrices enter into play
and hence these effects would vanish.

\subsection{Model Independent Expectations for LFV ? }
A crucial feature of the seesaw mechanism is that it has a larger number of 
parameters than those relevant for neutrino masses and
mixings. This would inhibit us in computing model independent
expectations for, say, BR($\mu \to e,\gamma$), given that the supersymmetric
seesaw mechanism works at the high scale. In fact, even after having a
complete knowledge of the entire neutrino mass matrix elements,
$\mathcal{M}_\nu$ as well as the heavy neutrino Majorana mass matrix
eigenvalues $M_{R_k}$, it is still not sufficient to completely
determine the rest of the seesaw parameters, namely the neutrino
Dirac Yukawa coupling matrix $h^\nu$.  This is best illustrated in the
parametrisation given by Casas and Ibarra \cite{casasibarra};
starting from the seesaw formula, Eq.~(\ref{seesaweq}), one can derive
$h^\nu$ in terms of low energy parameters as : 
\be
\label{rparameterisation}
h^\nu = U_{\rm PMNS}^\star ~\mathcal{D}_{\sqrt{\mathcal{M}_\nu}}~ R^T~ \mathcal{D}_{\sqrt{M_R}}~ 
{1 \over v_2},
\ee 
where $U_{\rm PMNS}$ is the Pontecorvo-Maki-Nakagawa-Sakata leptonic mixing matrix,
$~\mathcal{D}_{\sqrt{\mathcal{M}_\nu}}$ is the square root of the
diagonal matrix of light neutrino mass eigenvalues,
$\mathcal{D}_{\sqrt{M_R}}$ is the square root of the diagonal matrix
of heavy neutrino mass eigenvalues and $R$ is an arbitrary complex
orthogonal matrix such that $RR^T = 1$. The matrix $R$ parameterises
our ignorance of the neutrino Yukawas in spite of the complete
knowledge of the neutrino mass matrix.  Though it is difficult to give
a physical definition to `$R$'
it is important to note that
it can have physical consequences for lepton flavour violation (as
well as leptogenesis), even if the neutrino masses and mixings are
already accounted for.  For lepton flavour violation, the relevant
part of the seesaw parameter space are the entries in the matrix
$h^\nu h^{\nu~\dagger}$ which are now given by :
\be
\label{hnuhnudagrpara}
h^\nu~h^{\nu~\dagger} = U_{PMNS}^\star ~\mathcal{D}_{\sqrt{\mathcal{M}_\nu}}~ R^T~ 
\mathcal{D}_{M_R}~ R^\star~\mathcal{D}_{\sqrt{\mathcal{M}_\nu}}~U_{PMNS}^T~ {1 \over v_2^2}.
\ee 
$R$ can be parameterised by three angles and three phases.  Flavour
violation is now determined in terms of the angles and phases in $R$,
$U_{PMNS}$, $\mathcal{M}_\nu$ and $M_R$. For a given neutrino
spectrum, $U_{PMNS}$ and neutrino mass eigenvalues are (approximately)
known. However, the branching ratios can now be either enhanced or
suppressed depending on the parameters in $R$. Choosing $M_R$ to be
either completely hierarchical or degenerate leads to a reduction in
the number of parameters affecting the branching ratios. It is further
reduced if $R$ is chosen to be real. For example, if $M_R$ is
completely hierarchical and $R$ is real, basically only one angle would
affect the branching ratios. Similarly, when $M_R$ is degenerate and
$R$ is real, the branching ratios are independent of $R$. Lepton
flavour violating rates for various cases of interest have been
analysed in \cite{casasibarra}, special cases in \cite{kageyama1,bottumup}.

As mentioned earlier, flavour violation at the weak scale can be treated independently of the
source in terms of the mass insertion parameters $\Delta^l_{ij}$. In the seesaw model, these
$\Delta$ parameters are generated by RGE evolution and are thus proportional to the neutrino
Yukawa couplings. In mSUGRA, the relation between these two parameters is given as\footnote{Note that
only $LL$ type $\Delta$s are generated in the SUSY seesaw mechanism as long as one sticks to the
superpotential given in Eq.~(\ref{eq1}).} :
\be
\label{rgemi1}
(\Delta^l_{ij})_{\rm LL} \approx -{3 m_0^2+A_0^2 \over 8 \pi^2}~ C_{ij},
\ee
where $C_{ij}$ is defined to contain all the information from neutrino Yukawa couplings,
specifically, the left-mixing angles and the eigenvalues \cite{lavignac1,ellishisanoraidal}; 
it is given by :
\be
\label{lavignac}
C_{ij}  = [h^\nu h^{\nu~\dagger}]_{ij} \log{M_X \over M_R}.
\ee 
$C_{ij}$ is the part of the `seesaw' parameter space that can be probed by LFV experiments. 
The various upper bounds on $\delta^l_{ij}$ from the experimental limits on $l_j \to l_i, \gamma$
decay rates and other LFV processes can now be converted to upper bounds on off-diagonal $C_{ij}$ 
parameters \cite{lavignac1} for each point in the SUSY breaking parameter space. In Table I,
we present limits on $(\delta^l_{ij})_{\rm LL}$ from the present and upcoming limits on $l_j \to
l_i+\gamma$ processes \cite{ourprl,workinprogress}.

\begingroup
\begin{table}[ht]
\begin{tabular}{|c|c|c|}
\hline
~$(\delta^l)_{\rm LL}$~ & ~Present limits~ &~ Upcoming limits ~\\[0.4pt]
\hline
\hline
12 & $3 \times 10^{-4}$ & $8.5 \times 10^{-6}$ \\
13  & $0.09$ &  $0.02$  \\
23  &$0.09$ & $0.02 $ \\[0.4pt]
\hline
\hline
\end{tabular}
\caption{Limits on $\delta_{\rm LL}$ parameters derived in the mSUGRA model for tan $\beta$=10. 
The average slepton mass is chosen to be around 400 GeV and 
$m_{0}~ \sim~$ 400 GeV.
 Our results agree with those presented in Refs.~\cite{MS,HN}.}
\end{table}
\endgroup
The implied bounds on $C_{ij}$ can in turn be used to study constraints on neutrino Dirac Yukawa 
entries, for example, arising from a flavour model. The sensitivity of LFV experiments in probing 
the seesaw parameter space has been analysed by Ellis and collaborators \cite{ellishisanoraidal}.
Choosing  well defined points in supersymmetric parameter space, such as the popular 
Benchmark/Snowmass points, scatter plots of $C_{ij}$ parameter space\footnote{$C_{ij}$ is denoted
as $H_{ij}$ in Ref.~\cite{ellishisanoraidal}.} probed by several LFV processes such as
$\mu \to e, \gamma$ are presented. 

\subsection{CP violation in the lepton sector}
In addition to LFV, the SUSY seesaw can be responsible for several CP violating phenomena at
 both high and low energies in the leptonic sector. This can be facilitated by large amount of 
complex phases present in the seesaw couplings $h^\nu$ and/or $M_R$. We list some of these
phenomena below.  
(i) Leptogenesis \cite{buchreview} \\
(ii) CP violation in neutrino oscillations \cite{neutreview}\\
(iii) CP violation in $\Delta L=2$ processes, such as neutrinoless double beta decay
\cite{pascoli1} \\
(iv) CP violation in $\Delta L=1$ processes, such as
CP violation in $\mu \to 3 e$ etc \cite{okadacpv,elliscpv}. \\
(v) Leptonic EDMs \cite{ellisedm,masinaedm,farzanedm} \\
(vi) Slepton oscillations and CP violation in the sleptonic sector \cite{arkani}.\\
\noindent 
Since the source of both LFV and some of the above phenomena, say, leptogenesis, is the same set 
of parameters in the lagrangian, some amount of correlation can be expected between these phenomena 
for some regions of the entire seesaw parameter space. However, such a correlation is in general not 
guaranteed, even if the two phenomena exist simultaneously. Such a study was carried out, by random
scanning of the parameter space, by Ellis and collaborators in the case of hierarchical heavy neutrinos
in Ref.~\cite{ellisraidal1} and for degenerate heavy neutrinos in 
Ref.~\cite{ellisraidalyanagida1}.
Implications of low energy observables, specifically LFV, for leptogenesis and vice versa can also be 
studied by using the R-parametrisation introduced above for $h^\nu$,  or any of the other suitable 
parametrisations of a complex generic matrix, $h^\nu$. Such studies have been carried out in 
\cite{davidson123,joaquim,pascolirodej,petcovyag2}. 
Other related studies are \cite{ERY2,siyeon}. A detailed
analysis of correlations between all the leptonic phenomena is clearly beyond the scope of the present
work. 

\subsection{From LFV to seesaw parameters}
As we have seen so far, measuring the neutrino mass parameters $\Delta
m_{\rm atm}^2$, $\Delta m^2_\odot$ and the mixing angles $\theta_{ij}$ more
precisely would not lead us to any information on either the
left-mixing or the right-mixing angles present in $h^\nu$ or its
eigenvalues. While it also holds true within the Standard Model
seesaw, the major advantage in the supersymmetric seesaw is the rich
FCNC and CP violating phenomena associated with it. A natural question
that follows is: Can we determine all the seesaw parameters by
purely low energy experiments? At this juncture, frankly, the
question looks a bit ambitious given that we still have not yet
completed the determination of the neutrino mixing matrix angles and
phases as well as $sg(\Delta m^2_{\rm atm})$.  However, taking into
consideration that (i) we have the possibility of low energy
supersymmetry being observable at the LHC, (ii) being probed in great
detail at linear colliders, (iii) having improved sensitivity at the
upcoming facilities such as  MEGA and super-B/charm factories, and (iv)
improved determination of neutrino mass parameters at JPARC and long-base-line
experiments, such an analysis may be required in the near future.  As
we will discuss later on, assuming a `best case' scenario, the first
hint of SUSY seesaw might in fact come from a $\mu~ \to e, \gamma$
decay, probably even before the advent of the LHC
\cite{profumo}. Later,  more detailed evidence might pile up.
The authors of Ref.~\cite{davidsonibarra} have presented a 
discussion on evaluating the seesaw parameters from low energy
observables. It has been shown that although there is a one-to-one
correspondence between measurables at the weak-scale and the high-scale
 parameters, in practice, even in the simplest supersymmetric
theories, such an analysis could be formidable to achieve.  
This is particularly true regarding the additional phases present 
in the sneutrino/slepton mass matrices. 
However, we continue to remain optimistic and let the future decide.

\section{What could be the Neutrino Yukawa couplings  }
In the above, we have seen that due to our lack of knowledge
on the neutrino Yukawa couplings, our predictions for lepton flavour
violation are not effective enough in a purely bottom-up
approach. This can be considered as a limitation of the model we have
been working with, namely, the MSSM with right-handed neutrinos. A
possible way out then could be to enlarge the SM gauge group to a much
larger group, such as a Grand Unified Theory.  In these models,
typically quark and leptons sit in the same multiplets, leading to
relations between their Yukawa couplings. One would expect that
similar correlations would occur when the seesaw mechanism is
incorporated within a GUT. Under the simplest of the GUT groups,
`SU(5)', however, the right-handed neutrinos remain singlets and the
problem with LFV predictions still persists. One needs at least a
group encompassing the Pati--Salam ${\rm SU(4)_c}$ symmetry, like an SO(10)
model. Here we review one such example.

\subsection{GUT models: An SO(10) example}
In the $SO(10)$ gauge theory, all the known fermions and the 
right-handed neutrinos are unified in a single representation 
of the gauge group, the \textbf{16}. 
The product of two \textbf{16} matter representations can only couple to 
\textbf{10}, \textbf{120} or \textbf{126} representations, which can be 
formed by either a single Higgs field representation or a 
non-renormalisable product of representations of several Higgs fields.  
In either case, the  Yukawa matrices resulting from the couplings to 
\textbf{10} and \textbf{126} are complex-symmetric, whereas they are 
antisymmetric when the couplings are to the \textbf{120}. 
Thus, the most general $SO(10)$ superpotential relevant to
fermion masses can be written as
\be
W_{SO(10)} = h^{10}_{ij} 16_i~ 16_j~ 10 + h^{126}_{ij} 16_i~ 16_j~ 126 
+ h^{120}_{ij} 16_i~ 16_j~ 120,  
\ee
where $i,j$ refer to the generation indices. In terms of the SM fields, 
the Yukawa couplings relevant for fermion masses are given by \cite{strocchi}
\footnote{Recently, SO(10) couplings have also been evaluated for various renormalisable
and non-renormalisable couplings in \cite{so10couplings}.}:
\bea
16\ 16\ 10\ &\supset& 5\ ( u u^c + \nu \nu^c) + \bar 5\
(d d^c + e e^c), \nn\\
16\ 16\ 126\ &\supset& 1\ \nu^c \nu^c + 15\ \nu \nu +
5\ ( u u^c -3~ \nu \nu^c) + \bar{45}\ (d d^c -3~ e e^c), \nn\\
16\ 16\ 120\ &\supset& 5\ \nu \nu^c + 45\ u u^c +
\bar 5\ ( d d^c + e e^c) + \bar{45}\ (d d^c -3~ e e^c),
\label{su5content}
\eea
where we have specified the corresponding $SU(5)$  Higgs representations 
for each  of the couplings and all the fermions are left handed fields. 
The resulting mass matrices can be written as
\bea
\label{upmats}
M^u &= & M^5_{10} + M^5_{126} + M^{45}_{120}, \\
\label{numats}
M^\nu_{LR} &= & M^5_{10} - 3~ M^5_{126} + M^{5}_{120}, \\
\label{downmats}
M^d &= & M^{\bar{5}}_{10} +  M^{\bar{45}}_{126} + M^{\bar{5}}_{120} 
+ M^{\bar{45}}_{120}, \\
\label{clepmats}
M^e &= & M^{\bar{5}}_{10} - 3  M^{\bar{45}}_{126} + M^{\bar{5}}_{120} 
- 3 M^{\bar{45}}_{120}, \\
\label{lneutmats}
M^\nu_{LL} &=& M^{15}_{126}, \\
M^\nu_{R} &=& M^{1}_{126}.
\eea

A simple analysis of the above mass matrices leads us to the following
result:  \textit{At least one
of the Yukawa couplings in $h^\nu~ =~ v_u^{-1}~M^\nu_{LR}$ has 
to be as large as the top Yukawa coupling} \cite{oscar}. This result holds 
true in general, independently of
the choice of the Higgses responsible for the masses in
Eqs.~(\ref{upmats}), (\ref{numats}), provided that no accidental fine-tuned 
cancellations of the different contributions in Eq.~(\ref{numats}) are
present. If contributions from the \textbf{10}'s solely 
dominate, $h^\nu$ and $h^u$ would be equal.
If this occurs for the \textbf{126}'s, then $h^\nu =- 3~ h^u$ \cite{mohapsakita}. 
In case both of them have dominant
entries, barring a rather precisely fine-tuned
cancellation between $M^5_{10}$ and $M^5_{126}$ in
Eq.~(\ref{numats}), we expect at least one large entry to be present in $h^\nu$.
A dominant antisymmetric contribution to top quark mass 
due to the {\bf 120} Higgs is phenomenologically excluded, since it would  
lead to at least a pair of heavy degenerate up quarks. 

Apart from sharing the property that at least one eigenvalue of both $M^u$ 
and $M^\nu_{LR}$ has to be large, for the rest it is clear from 
(\ref{upmats}) and  (\ref{numats}) that these two matrices are not aligned 
in general, and hence we may expect different mixing angles appearing 
from their diagonalisation. This freedom is removed if one sticks to 
particularly simple choices of the Higgses responsible for up quark and
neutrino masses. A couple of remarks are in order here. Firstly, note that
in general there can be an additional contribution, Eq.~(\ref{lneutmats}),
to the light neutrino mass matrix, independent of the canonical seesaw mechanism. Taking 
into consideration also this contribution leads to the so-called Type-II seesaw 
formula \cite{type2seesaw}. Secondly, the correlation between neutrino Dirac Yukawa 
coupling and the top Yukawa is in general independent of the type of seesaw mechanism, 
and thus holds true irrespective  of the light-neutrino mass structure. 

\subsection{What the neutrino Yukawa mixing matrices could be}
Within the $SO(10)$ example, we have seen above that the amount of
mixing present in the neutrino Dirac Yukawa couplings, $h^\nu$ depends
on the type and number of Higgs representations there are in the
theory. Motivated by the flavour structure of Standard Model fermions
(including neutrino mixing angles), one can imagine two cases of
mixings to be present in $h^\nu$.  The first one corresponds to a case
where the mixing present in $h^\nu$ is small and CKM-like.  We will
call this case, `the minimal case'. This is typical of models in which
quarks and leptons have the same mixing angles at the high
scale.  The required large mixing angles in the light-neutrino sector
is purely resultant of the seesaw mechanism. As a second case, we
consider scenarios where the mixing in $h^\nu$ is no longer small, but
large like the observed PMNS mixing. In this case, the heavy neutrino
mass matrix only plays the role of a large scale due to which light
neutrino masses are suppressed. We will call this case the `the
maximal case'
\footnote{Note that these two cases are the two examples of how one
can generate large mixing angles for light-neutrino mass matrices from
the seesaw mechanism \cite{seesawreviews}.}. These two cases
serve as `benchmark' scenarios for seesaw induced lepton flavour
violation in SUSY $SO(10)$.  Similar studies of these two extreme
cases have also been considered in \cite{hisano,ellishisanoraidal}.

\subsubsection{Minimal case: A model for CKM mixings} 

The minimal Higgs spectrum to obtain phenomenologically viable mass matrices
includes two \textbf{10}-plets, one coupling to
the up-sector and the other to the down-sector. In this way it is possible to 
obtain the required CKM mixing \cite{buchwyler} in the quark sector. 
The $SO(10)$
superpotential is now given by

\be
\label{primedbasis}
W_{SO(10)} = {1 \over 2}~  h^{u,\nu}_{ij} 16_i~ 16_j~ 10_u + 
{1 \over 2}~ h^{d,e}_{ij} 16_i ~16_j~ 10_d + {1 \over 2}~ h^R_{ij}~ 16_i~ 16_j~ 126. 
\ee
We further assume that the {\bf 126} dimensional Higgs field gives
Majorana mass  \textit{only} to the right-handed neutrinos. 
An additional feature of the above mass matrices is that all of them 
are \textit{symmetric}. From here, it is clear that the following mass 
relations hold between the quark and leptonic mass matrices at the GUT 
scale\footnote{Clearly this relation cannot hold for the first two 
generations of down quarks and charged leptons.  One expects, small 
corrections due to  non-renormalisable operators or
suppressed renormalisable operators \cite{georgijarlskog} to be invoked.}: 
\be
\label{massrelations}
h^u  = h^\nu \;\;\;;\;\;\; h^d  = h^e . 
\ee
In the basis where charged lepton masses are diagonal, we have
\be
\label{hnumg}
 h^\nu = V_{\rm CKM}^T~ h^u_{diag}~ V_{\rm CKM}. 
\ee 
The large couplings in $h^\nu \sim {\mathcal O}(h_t)$ induce 
significant off-diagonal entries in $m_{\tilde L}^2$ through the RG
evolution between $M_{\rm GUT}$ and the scale of the right-handed Majorana neutrinos
\footnote{Typically one has different mass scales associated with different
right-handed neutrino masses.}, $M_{R_i}$. The induced off-diagonal entries
relevant to $l_j \rightarrow l_i, \gamma$ are of the order of: 
\noindent 
\begin{eqnarray}
\label{wcmi1}
(m_{\tilde L}^2)_{21}&\approx& -{3 m_0^2+A_0^2 \over 8 \pi^2}~
h_t^2 V_{td} V_{ts} \ln{M_{\rm GUT} \over M_{R_3}} 
+ \mathcal{O}(h_c^2), \\
\label{wcmi2}
(m_{\tilde L}^2)_{32}&\approx& -{3 m_0^2+A_0^2 \over 8 \pi^2}~
h_t^2 V_{tb} V_{ts} \ln{M_{\rm GUT} \over M_{R_3}} 
+ \mathcal{O}(h_c^2), \\
\label{wcmi3}
(m_{\tilde L}^2)_{31}&\approx& -{3 m_0^2+A_0^2 \over 8 \pi^2}~
h_t^2 V_{tb} V_{td} \ln{M_{\rm GUT} \over M_{R_3}} 
+ \mathcal{O}(h_c^2).
\end{eqnarray}
\noindent 
In these expressions, the CKM angles are small but one would expect 
the presence of the large top Yukawa coupling to compensate such a 
suppression. The required right-handed neutrino Majorana  mass 
matrix, consistent with both the observed low energy neutrino masses 
and mixings as well as with CKM-like mixings in $h^\nu$ is easily determined 
from the seesaw formula defined at the scale of right-handed neutrinos
\footnote{ The neutrino masses and mixings here are defined
at $M_{R}$. Radiative corrections can significantly modify the neutrino
spectrum from that of the weak scale \cite{nurad}. This is more true for 
the degenerate spectrum of neutrino masses \cite{degenrad} and for some 
specific forms of $h^\nu$ \cite{antusch}. For our present discussion, with
hierarchical neutrino masses and up-quark like neutrino Yukawa matrices, we 
expect these effects not to play a very significant role.}  
\be
\label{MRstrcture1}
M_R = V_{\rm CKM}~ h_{diag}^u~ V_{\rm CKM}^T~ m_{\nu}^{-1} V_{\rm CKM}~ h_{diag}^u~ V_{\rm CKM}^T, 
\ee   
where we have used Eq.~(\ref{hnumg}) for $h^\nu$.
For hierarchical neutrino mass spectrum, $m_{\nu_3} \approx \sqrt{\Delta m^2_{\rm atm}}$,
$m_{\nu_{2}} \approx \sqrt{\Delta m^2_{\odot}}$, and 
$m_{\nu_{1}} \ll \sqrt{\Delta m^2_{\odot}}$ and for a nearly bi-maximal
$U_{\rm PMNS}$, 
it is straightforward to see that  the right handed neutrino mass 
eigenvalues are given by:
\be
\label{mrapproxegckm}
M_{R_3} \approx {m_t^2 \over 4~ m_{\nu_1}} \;;\;
M_{R_2} \approx {m_c^2 \over 4~ m_{\nu_1}} \;;\;
M_{R_1} \approx {m_u^2 \over 2~ m_{\nu_1}} \;.
\ee 

\begin{figure}[ht]
\includegraphics[scale=0.35]{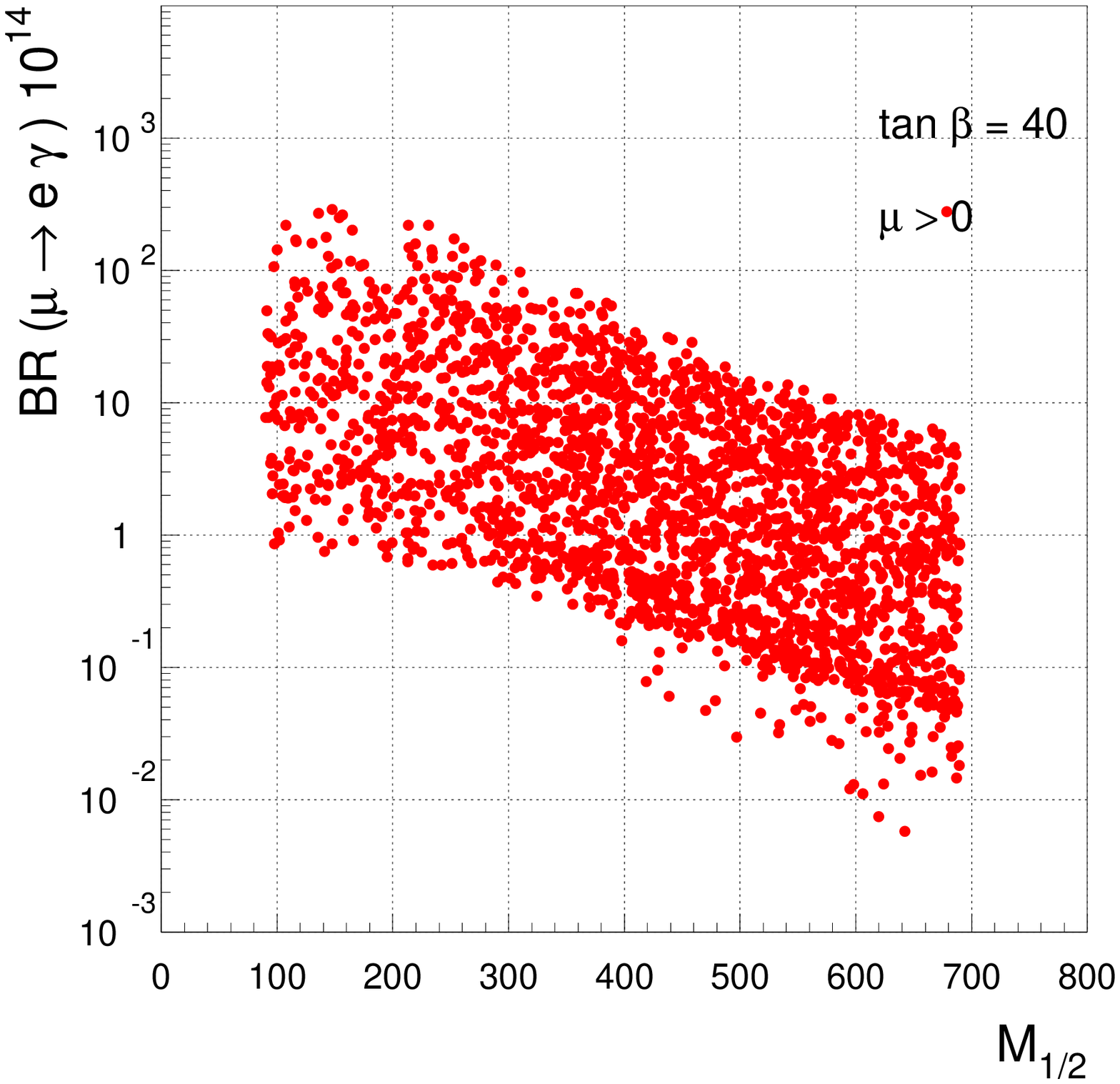}
\caption{The scatter plots of branching ratios of $\mu \to e, \gamma$
decays as a function of $M_{1/2}$ are shown for the (minimal) CKM case for 
tan $\beta$ = 40. Results do not alter significantly with the change 
of sign($\mu$).}
\end{figure}

The Br($\mu \to e , \gamma$) is now predictable in this case. Considering
mSUGRA boundary conditions, we compute these branching ratios numerically. 
In Fig. 2) we show the scatter plots (in mSUGRA parameter space $m_0,~A_0,~M_{1/2}$) 
for BR($\mu \rightarrow e,\gamma$ ) for the CKM case 
and $\tan \beta = 40$. We see that 
reaching a sensitivity of $10^{-14}$ for 
BR$(\mu \to e \gamma)$ would allow us to probe the SUSY 
spectrum  completely up to $M_{1/2} = 300$ GeV (notice that this corresponds to 
gluino and squark masses of order 750 GeV) and would still probe 
large regions of the parameter space up to $M_{1/2} = 700$ GeV.  
Thus, in summary, though the present limits on BR($\mu \to e, \gamma$) 
would not induce any significant constraints on the supersymmetry-breaking
parameter space, an improvement in the limit to $\sim {\mathcal O }(10^{-14})$,
as foreseen, would start imposing non-trivial constraints especially
for the large $\tan \beta$ region. 

A further comment on the `minimal' mixing case is in order. Strictly
speaking this is not the `minimalest' mixing possible in the Dirac neutrino Yukawa
couplings. It has been shown in models where the right-handed neutrinos attain
their masses through Yukawa couplings, one can essentially set the Dirac neutrino 
Yukawa mixing to be zero and the entire mixing comes from the right handed neutrino
sector \cite{babumohapdutta1}. Such a situation can be realised within left-right
symmetric models, with or without an SO(10) embedding. The renormalisation group
flow is different in this case and the LFV is now related to Yukawa couplings of
the right-handed neutrino mass matrix in an indirect manner \cite{babumohapdutta1,mohapdutta1}.

\subsubsection{Maximal case: A way for PMNS mixings}

The minimal $SO(10)$ model presented in the previous subsection would
inevitably lead to small mixing in $h^\nu$. In fact, with two Higgs fields
in symmetric representations, giving masses to the up-sector and the 
down-sector separately, it would be difficult to avoid the small CKM-like
mixing in $h^\nu$. To generate mixing angles larger than CKM angles, asymmetric
mass matrices have to be considered. In general, it is sufficient to introduce
asymmetric textures either in the up-sector or in the down-sector. In the 
present case, we assume that the down-sector couples to a combination of Higgs 
representations (symmetric and antisymmetric)\footnote{The couplings of $\Phi$ 
in the superpotential can be either renormalisable or non-renormalisable. 
See \cite{chang} for a non-renormalisable example.}
$\Phi$, leading to an asymmetric mass matrix in the basis where the up-sector
is diagonal. As we will see below, this would also require that the right-handed
Majorana mass matrix be diagonal in this basis. We have : 

\be
\label{mnsso10}
W_{SO(10)} = {1 \over 2}~  h^{u,\nu}_{ii}~ 16_i ~16_i 10^u + 
{1 \over 2}~ h^{d,e}_{ij}~ 16_i ~16_j \Phi \nn \\
 + {1 \over 2}~ h^R_{ii}~ 16_i~ 16_i 126~, 
\ee
where the \textbf{126}, as before, generates only the right-handed neutrino
mass matrix. To study the consequences
of these assumptions, we see that at the level of $SU(5)$, we have
\be
W_{SU(5)} = {1 \over 2}~ h^u_{ii}~ 10_i ~10_i ~5_u 
+ h^\nu_{ii} ~\bar{5}_i~ 1_i~ 5_u \nn \\
+  h^d_{ij}~ 10_i ~\bar{5}_j~ \bar{5}_d 
+  {1 \over 2}~M^R_{ii}~ 1_i 1_i,
\ee
where we have decomposed the $16$ into $10 + \bar{5} + 1$ and $5_u$ and
$\bar{5}_d$ are components of $10_u$ and $\Phi$ respectively. To have large
mixing $\sim~ U_{\rm PMNS}$  in $h^\nu$ we see that the asymmetric matrix $h^d$
should now be able to generate both the CKM mixing as well as PMNS mixing. 
This is possible if 
\be
V_{\rm CKM}^T~ h^d~ U_{\rm PMNS}^T = h^d_{diag}. 
\ee 

\begin{figure}[ht]
\includegraphics[scale=0.35]{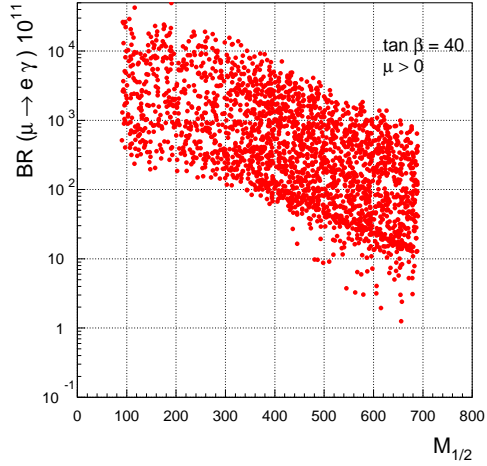}
\caption{The scatter plots of branching ratios of $\mu \to e, \gamma$
decays as a function of $M_{1/2}$ are shown for the (maximal) PMNS case for 
tan $\beta$ = 40. The results do not alter significantly with the change 
of sign($\mu$).}
\end{figure}

This would mean that the $10$ that contains the left-handed down-quarks would
be rotated by the CKM matrix whereas the $\bar{5}$ that contains the left
handed charged leptons would  be rotated by the $U_{\rm PMNS}$ matrix to go into
their respective mass bases \cite{moroi,chang}. Thus we have, in analogy with the
previous subsection, the following relations to hold true in the basis where 
charged leptons and down quarks are diagonal:
\bea
h^u &=& V_{\rm CKM}~ h^u_{diag}~ V_{\rm CKM}^T~ ,  \\
\label{hnumns}
h^\nu &=& U_{\rm PMNS}~ h^u_{diag}. 
\eea
Using the seesaw formula of Eqs.~({\ref{seesaweq}) and (\ref{hnumns}), we have
\be
M_{R} = diag\{ {m_u^2 \over m_{\nu_1}},~{m_c^2 \over m_{\nu_2}},
~{m_t^2 \over m_{\nu_3}} \}. 
\ee 
We now turn our attention to  lepton flavour violation in this case. The
branching ratio, BR($\mu \rightarrow e, \gamma)$ would now be dependent on: 
\be
\label{hnusqmns}
[h^\nu h^{\nu~T}]_{21} = h_t^2~ U_{\mu 3}~ U_{e 3} + h_c^2~ U_{\mu 2}~ U_{e 2} +
\mathcal{O}(h_u^2). 
\ee 
It is clear from the above that in contrast to the CKM case,
the dominant contribution to the off-diagonal entries depends on the 
unknown magnitude of the element $U_{e3}$ \cite{ue3importance}. 
If $U_{e3}$ is very close to its
present limit $\sim~0.2$ \cite{chooz}, the first term on the 
RHS of the Eq.~(\ref{hnusqmns})
would dominate. Moreover, this would lead to large contributions to the 
off-diagonal entries in the slepton masses with $U_{\mu 3}$ of 
${\mathcal O}(1)$. We have :
\be
\label{bcmi1}
(m_{\tilde L}^2)_{21} \approx -{3 m_0^2+A_0^2 \over 8 \pi^2}~
h_t^2 U_{e 3} U_{\mu 3} \ln{M_{\rm GUT} \over M_{R_3}}
+ \mathcal{O}(h_c^2). 
\ee
The above contribution is larger than the CKM case by a factor of 
$(U_{\mu 3} U_{e3})/ (V_{td} V_{ts}) \sim 140 $ compared with the CKM case. 
From Eq.~(\ref{BR}) we see that 
it would mean about a factor $10^4$ times larger than the CKM case in 
BR($\mu \rightarrow e, \gamma)$. In case $U_{e3}$ is very small, \textit{i.e} 
either zero or $\ler~ (h_c^2/h_t^2)~U_{e2}~ \sim 4 \times 10^{-5}$, the 
second term $\propto ~h_c^2$
in Eq.~(\ref{hnusqmns}) would dominate. However the off-diagonal contribution
in slepton masses, now being proportional to charm Yukawa could be much smaller, 
in fact even  smaller than the CKM contribution by a factor 
\be{h_c^2~ U_{\mu 2} ~U_{e 2} \over  h_t^2 ~V_{td} ~V_{ts}} 
\sim  7 \times 10^{-2}.
\ee
If $U_{e3}$ is close to its present limit, the current bound on
BR($\mu \rightarrow e, \gamma$) would already be sufficient to
produce stringent limits on the SUSY mass spectrum. Similar
$U_{e3}$ dependence can be expected in the $\tau \to e $ transitions
where the off-diagonal entries are given by : 
\be
\label{bcmi3}
(m_{\tilde L}^2)_{31} \approx  -{3 m_0^2+A_0^2 \over 8 \pi^2}~
h_t^2 U_{e 3} U_{\tau 3} \ln{M_{\rm GUT} \over M_{R_3}}
+ \mathcal{O}(h_c^2). 
\ee
The $\tau \to \mu$ transitions are instead $U_{e3}$-independent probes of SUSY,
whose importance was first pointed out in Ref.~\cite{kingblazektmg}. As in the
rest of the cases, the off-diagonal entry in this case is given by :
\be
\label{bcmi2}
(m_{\tilde L}^2)_{32} \approx  -{3 m_0^2+A_0^2 \over 8 \pi^2}~
h_t^2 U_{\mu 3} U_{\tau 3} \ln{M_{\rm GUT} \over M_{R_3}} 
+ \mathcal{O}(h_c^2). 
\ee

In the PMNS scenario, Fig. 3 shows the plot for BR($\mu
\rightarrow e, \gamma$) for tan $\beta$ = 40. In this plot, the value of
$U_{e3}$ chosen is very close to the present experimental upper limit
\cite{chooz}. As long as $U_{e3} \ger 4 \times 10^{-5}$, the plots
scale as $U_{e3}^2$, while for $U_{e3} \ler 4 \times 10^{-5}$ the term
proportional to $m_c^2$ in Eq.~(\ref{bcmi1}) starts dominating; 
the result is then insensitive to the choice of $U_{e3}$.  
For instance, a value of $U_{e3} =0.01$ would reduce the BR by a factor of
225 and still a significant amount of  the parameter space 
for $\tan \beta = 40$ would
be excluded. 
We further find that with the present limit on BR($\mu \rightarrow e, \gamma$),
all the parameter space would be completely excluded up to $M_{1/2}=
300$ GeV for $U_{e3} =0.15$, for any vale of $\tan \beta$ (not shown in the
figure).

In the $\tau \to \mu \gamma$ decay the situation is similarly
constrained.  For $\tan \beta=2$, the present bound of $3 \times
10^{-7}$ starts probing the parameter space up to $M_{1/2} \leq 150$
GeV. The main difference is that this does not depend on the value of
$U_{e3}$, and therefore it is already a very important constraint on
the parameter space of the model.  In fact, for large $\tan \beta =
40$, as shown in Fig. 4, reaching the expected limit of $1 \times
10^{-8}$ would be able to rule out completely this scenario up to
gaugino masses of $400$ GeV, and only a small portion of the parameter
space with heavier gauginos would survive. In the limit $U_{e3} =0$,
this decay mode would provide a constraint on the model stronger than
$\mu \to e, \gamma$, which would now be suppressed as it would contain
only contributions proportional to $h_c^2$, as shown in
Eq.~(\ref{bcmi1}).

\begin{figure}[ht]
\includegraphics[scale=0.35]{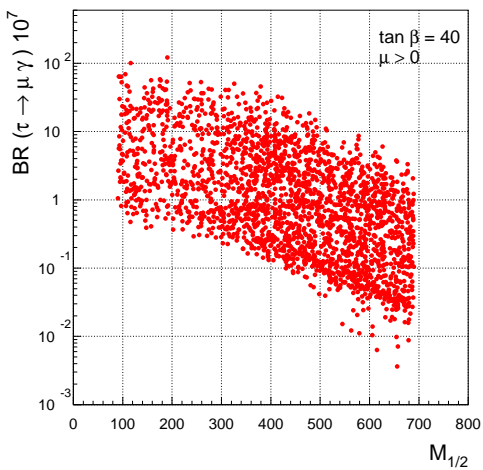}
\caption{The scatter plots of branching ratios of $\tau \to \mu, \gamma$
decays as a function of $M_{1/2}$ are shown for the (maximal) PMNS case for 
tan $\beta$ = 40. The results do not alter significantly with the change 
of sign($\mu$).}
\end{figure}

In summary, in the PMNS/maximal mixing case, even the present limits
from BR($\mu \to e, \gamma$ ) can rule out large portions of the supersymmetry-breaking 
parameter space, if $U_{e3}$ is either close to its present limit
or within an order of magnitude of it (as the planned experiments
might find out soon \cite{goodman}). These are more severe for the large $\tan \beta$
case. In the extreme situation of $U_{e3}$ being zero or very small $\sim 
{\mathcal O}(10^{-4} - 10^{-5})$, BR($\tau \to \mu, \gamma$) will start
playing an important role, with its present constraints already disallowing
large regions of the parameter space at large $\tan \beta$. 
While the above example concentrated on the hierarchical light neutrinos, 
similar `benchmark' mixing scenarios have been explored in great detail, for degenerate 
spectra of light neutrinos, by Ref.~\cite{illanamasip}, taking
also in to consideration running between the Planck scale and the GUT scale. 

\subsection{Textures and other examples}
While the bottom-up approach studies the possibility of  `measuring' the 
seesaw parameters through low energy experiments, the top-down approach 
gives the opportunity to study several theoretically well motivated models
encompassing the seesaw mechanism. For example, a flavour symmetry based on either 
abelian or non-abelian family symmetries could be at work at
the scales giving rise to specific patterns in all the Yukawa coupling matrices
in the lagrangian including that of neutrino Dirac Yukawa couplings. Low energy
flavour violation is then dependent on these patterns of the Yukawa matrices 
which are predictable. Several analyses of this kind have been presented in the
literature mostly within a supersymmetric 
GUT \cite{lavignac1,textures,kingpeddie,lopsidedlfv,mahanchen1}. A recent review of several
textures presented in the literature can be found in \cite{mahanchen2}.

Textures that tend to lead to large left-mixing in $h^\nu$ are typically prone to 
constraints from the present limits on $\mu \to e + \gamma$. A class of textures which 
goes by the name of `lop-sided' generically predict large branching ratios \cite{lopsidedlfv} 
within the reach of experiments at MEGA. Finally, in addition to the $SO(10)$ example
 presented here, there have been several other examples both within the context 
of $SO(10)$ and otherwise that have been explored in the 
literature \cite{topdown1,patiso10}.

\section{Seesaw induced LFV and associated Phenomenology }
So far we have looked at LFV generated by a seesaw mechanism both
through a bottom-up as well as top-down perspectives. In the following
we will discuss the impact the generated LFV can have on the
associated phenomenology of the SUSY model. We will only concentrate on
a few issues, leaving out the leptonic CP violation which we have
already commented on previously. Most of these correlated effects are
only valid within a class of SUSY-GUT models, as such correlations
cannot be constructed from a purely bottom-up approach.

\subsection{LFV and experimental sensitivity to $U_{e3}$ }

\begin{figure}[ht]
\label{ue3}
\includegraphics[scale=0.40,angle=-90]{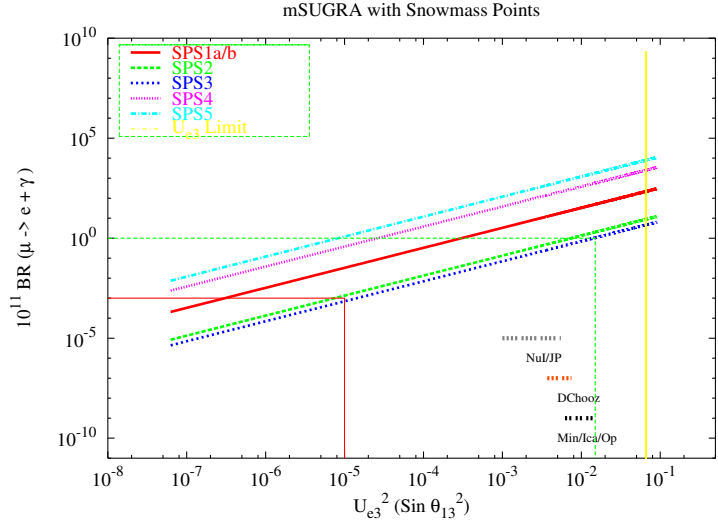}
\caption{The variation of the BR($\mu \to e + \gamma$) with respect to $U_{e3}^2$ in 
the maximal mixing, MNS case. Each (diagonal) line corresponds to an SPS point in mSUGRA 
 as denoted in the legend in the figure. The sensitivity of future LFV experiments ~ 
$\mathcal{O}$($10^{-11}$--$10^{-14}$) is projected on the $U_{e3}^2$ axis. 
The ranges probed by future long-base-line experiments are shown
as horizontal lines. NuI/JP represents he projected sensitivity of Nu-MI/J-PARC, DChooz, 
reactor-based experiments like  double CHOOZ, Min/Ica/Op represents the experiments MINOS,
ICARUS and OPERA \cite{lindner}.}
\end{figure}

In the maximal mixing situation, which we have discussed in the previous subsection, we
have seen that the BR($\mu \to e, \gamma$) would depend crucially on the neutrino mixing
matrix element $U_{e3}$.
To illustrate the correlations between $\mu \to e + \gamma$ and the neutrino 
mixing angle $U_{e3}$, we chose specific points in the supersymmetric parameter
space as given by the Snowmass collaboration \cite{snowmass}. These are presented in
Table II. 

\begin{table}
\begin{tabular}{|c|c|c|c|c|c|}
\hline
 &~ $m_0$~ &~ $M_{1/2}$~ &~ $A_0$~ &~ tan $\beta$~ & ~sg($\mu$)~\\[0.12pt]
\hline
~SPS1a~&~100~&~250~&~$-$100~&~10&~$>0$~\\
~SPS1b~&~200~&~400~&~0~&~30~&~$>0$~\\
~SPS2~&~1450~&~300~&~0~&~10~&~$>0$~\\
~SPS3~&~90~&~400~&~0~&~10~&~$>0$~\\
~SPS4~&~400~&~300~&~0~&~50~&~$>0$~\\
~SPS5~&~150~&~300~&~$-$1000~&~5~&~$>0$~ \\[0.12pt]
\hline
\end{tabular}
\caption{SPS points for mSUGRA }
\end{table} 

In Fig. 2, we plot BR($\mu \to e + \gamma$) with respect to $U_{e3}^2$ or
$\sin^2 \theta_{13}$. We also present the sensitivity of various future experiments
probing $U_{e3}$ and well as the expected improvements on the limits on 
BR($\mu \to e + \gamma$). It is seen that $\mu \to e \gamma$ can
have a stronger sensitivity on $U_{e3}$ if both SUSY seesaw and maximal mixing 
case are realised in nature \cite{neut04}. A detailed study of the impact of 
$U_{e3}$ on observability of SUSY at the LHC can be found in \cite{profumo}.

\subsection{Correlations with other SUSY search strategies}
In addition to the improvements in LFV experiments, this is also 
going to be the decade in which we should be able to establish whether
low energy supersymmetry exists or not through direct searches
at the LHC \cite{lhc}. On the other hand, 
improved astrophysical observations from experiments like 
WMAP \cite{wmap} and Planck are going to determine the relic density 
of supersymmetric LSP at unprecedented accuracy. Within mSUGRA, 
correlations between these two search strategies have been studied
\cite{baertata}. Incorporating the seesaw mechanism in the model 
\`a la SO(10), would generate another discovery strategy through
the lepton flavour violation channel.  This is especially
true when the LFV entries in the slepton mass matrices are maximised,
as in the PMNS case. 

We see that three main regions in the mSUGRA parameter space would 
survive after imposing
all the present phenomenological and astrophysical (dark matter)
constraints \cite{profumo}\footnote{For a bottom-up analyses,
 see Ref.~\cite{campbell}.}.
These are: (a) The stau coannihilation regions, where the lightest stau
is quasi-degenerate with the neutralino LSP and efficient stau-stau as well
as stau-neutralino (co)annihilations suppress the relic density. (b) The
A-pole funnel region, where the neutralino(bino)-neutralino annihilation
process is greatly enhanced through a resonant s-channel exchange of the heavy
neutral Higgs A and H (c) Focus point or hyperbolic branch regions, where a
non-negligible higgsino fraction in the lightest neutralino is produced. 
In each of these regions the LFV rates emanating from the seesaw mechanism 
can be computed and contrasted with the sensitivity of direct searches at the
LHC. Assuming the maximal mixing PMNS case, we find \cite{profumo}:

\begin{itemize}

\item \textit{Coannihilation regions}: In these regions, which are
mostly accessible at the LHC, an improvement of two orders of magnitude in 
the branching ratio sensitivity from the present limit, 
would make $\mu \to e \gamma$ visible 
for most of the parameter space, as long as $U_{e3} \gtrsim 0.02$, even for 
the low tan $\beta$ region. For large tan $\beta$, independent of $U_{e3}$,
$\tau \to \mu \gamma$ will start probing this region provided a sensitivity
of $\mathcal{O}(10^{-8})$ is reached.  

\item \textit{A-pole funnel regions}: In these regions the LHC
reach is not complete and LFV may be competitive. If $U_{e3} \gtrsim 10^{-2}$,
the future $\mu \to e \gamma$ experiments, with limit of 
${\mathcal O}(10^{-14})$ will probe most of the parameter 
space. 
As before, $\tau \to \mu \gamma$ will probe this region once the BR 
sensitivity reaches $\mathcal{O}(10^{-8})$.

\item \textit{Focus point regions}: Since the LHC reach in this region is 
rather limited by the
large $m_0$ and $M_{1/2}$ values, LFV could constitute a privileged road 
toward SUSY discovery. This would require improvements of at least a couple
of orders of magnitude (or more, depending on the value of $U_{e3}$) of
improvement on the present limit of BR($\mu \to e, \gamma$). 
Dark Matter (DM) searches will also have partial access to this region in future, 
leading to a new complementarity between LFV and the quest for the cold dark matter
 constituent of the Universe.
\end{itemize}

\subsection{Seesaw induced Hadronic FCNC and CPV}
So far we have seen that the SUSY version of the seesaw mechanism can 
lead to potentially large leptonic flavour violations, so large that they
 could compete even with  direct searches at the LHC. If one combines 
these ideas of supersymmetric seesaw with those of quark--lepton unification,
 as in a supersymmetric GUT, one would expect that
 the seesaw resultant flavour effects would now be also felt in the
 hadronic sector, and vice versa \cite{hkr,bhs}. In fact, this is what
 happens in a SUSY SU(5) with seesaw mechanism \cite{moroi}, where the
 seesaw induced  RGE effects generate flavour violating terms in 
the right handed squark multiplets. 
However, as is the case with the MSSM + seesaw mechanism, within the SU(5)
model also, information from the neutrino masses is not sufficient to 
fix all the seesaw parameters; a large neutrino Yukawa coupling has
to be \textit{assumed} to have the relevant phenomenological consequences
in hadronic physics, such as CP violation in $B \to \Phi K_s$ etc. 

As we have already seen within the SO(10) model, a large neutrino
Yukawa, of the order of that of the  top quark, is almost inevitable. 
Using this, it has been pointed in Ref.~\cite{chang}, that the observed
large atmospheric $\nu_\mu$ -$\nu_\tau$ transitions imply a potentially
large $b \to s$ transitions in SUSY SO(10). In the presence of CP
violating phases, this can lead to enhanced CP asymmetries in $B_s$ and $B_d$
decays. In particular, the still controversial discrepancy between the 
SM prediction and the observed $A_{\rm CP}(B_d \to \Phi K_s)$ \cite{bellebabar}
 can be attributed
to these effects. Interestingly, despite the severe constraints on the
$b \to s$ transitions from $B \to X_s,\gamma$ \cite{bsgamma,borzumati},
 subsequent detailed analyses
\cite{otherbs,rome} proved that there is still enough room for sizeble
deviations from the SM expectations for CP violation in the $B$ systems. 
The reader interested in various correlations in $b \to s$ transitions
with all possible FV off-diagonal squark mass entries can find an exhaustive
answer in Ref.~\cite{rome}. 

Finally, let us make a short comment about possible correlations between 
the hadronic and leptonic FV effects in a SUSY GUT. If the FV soft 
breaking terms appear at a scale larger than that of the grand unification, 
they must be related by the GUT symmetry. This puts constraints on the
boundary conditions for the running of the FV soft parameters. From this
consideration, one might intuitively expect that some correlation between 
various leptonic and hadronic FCNC processes \cite{ourprl} can occur at the weak 
scale. If in the evolution of the sparticle masses from the grand unification
scale down to the electroweak scale, one encounters seesaw physics, then
the quark--lepton correlations involving the left-handed sleptons, though 
modified, lead to even stronger constraints on hadronic
 physics \cite{ourprl,workinprogress}. Some other related works are \cite{hisanoedm}.

\section{Alternatives to Canonical Seesaw and LFV }
In the final section before conclusions, we briefly mention lepton flavour
violating studies conducted in mechanisms other than the canonical seesaw. 
As earlier, our list is not exhaustive or detailed. \\
\noindent
(i) \textit{Type II seesaw}: In the $SO(10)$ example, as we have shown, 
one can consider a situation where the non-seesaw contribution (\ref{lneutmats}) dominates
over the seesaw one. Flavour violation in this case has been studied in 
specific models \cite{mohaptype2}.  \\
(ii) \textit{Triplet Seesaw}: The seesaw mechanism itself can be implemented with fermionic
triplets instead of singlet fermions. In the supersymmetric context, the flavour violation
is then proportional to the triplet couplings and the neutrino mixing angles \cite{tripletseesaw}.\\
(iii) \textit{3 $\times$ 2  Seesaw}: The seesaw parameter space drastically reduces when
one of the neutrinos is assumed to be too heavy and decouples from the model. The predictive
power of the model is now enhanced for various low energy and high energy 
observables \cite{ross3by2,pokorski3by2,mohap3by2}.  \\
(iv) \textit{X-dimensional models}: In models with warped X-dimensional scenarios, large
LFV is expected with either Dirac-type neutrinos \cite{xdims1} or Majorana-type neutrinos \cite{xdims2}.\\
Flavour violation has also been computed in various other models as R-parity violation \cite{romao},
seesaw models with additional leptons (or inverse seesaw) \cite{valle}, left--right symmetric
models \cite{vogel}, Standard Model seesaw with a large number of Higgs bosons \cite{grimus}. 

\section{Conclusions}
Since the discovery of neutrino masses in atmospheric neutrino oscillations, there has been a lot
of activity trying to understand the seesaw mechanism and its low energy implications. On the
one side, we use a bottom-up approach quantifying our ignorance on the seesaw parameter space to
compute the LFV, while on the other side, in top-down approach, various models and textures 
are constrained by LFV. Associated implications to $B$-physics GUT models, 
DM abundances in SUGRA theories and implications for LHC searches have been and are
being done. Though we have not discussed it here, 
LFV in $Z$-decays \cite{illana} and other 
collider processes \cite{ruckl} is also being investigated. 

Undoubtedly, the seesaw mechanism represents (one of ) the best 
proposal to generate small neutrino masses naturally. But, how
can we make sure that this is indeed nature's choice ? 
Even establishing the Majorana nature of the neutrinos through a
positive evidence of neutrinoless double beta decay, it will be
difficult to assess that such Majorana  masses come from a seesaw. 
Indeed, as we said at the beginning, in the SM seesaw we expect 
very tiny charged LFV effects, probably without any chance of ever
observing them. When moving to SUSY seesaw we add an important handle
to our effort to establish the presence of a seesaw. In fact, as we
tried to show in this work, SUSY extensions of the SM with a seesaw
have a general `tendency' to enhance (or even strongly enhance) rare
LFV processes. Hence the combination of the observation of neutrinoless
double beta decay and of some charged LFV phenomenon would constitute
an important clue for the assessment of SUSY seesaw in nature.

There is no doubt that after the discovery of the neutrino masses, 
among the indirect tests of SUSY through FCNC and CP violating 
phenomena, LFV processes have acquired a position of utmost relevance. 
It would be spectacular if, by the time the LHC observes the first SUSY
particle, we could see also a muon decaying to an electron and a photon! 
After thirty years, we could have the simultaneous confirmation of two
of the most challenging physics ideas: seesaw and low energy SUSY. 

\textbf{Acknowledgements:} We thank all our collaborators, Marco Ciuchini, 
Stefano Profumo, Luca Silvestrini and Carlos Yaguna, without whom the present
work would not have been possible. Diagram by Jaxodraw \cite{jax}.

\end{document}